# Power Loading based on Portfolio Theory for Densified Millimeter-Wave Small-Cell Communications

Shuyi Shen[1], Bernardo A. Huberman[2], Lin Cheng[2], and Gee-Kung Chang[1]
*1. School of Electrical and Computer Engineering, Georgia Institute of Technology, Atlanta, GA, USA;*
*2. CableLabs, Louisville, CO, USA.*
*ssyzoe@gatech.edu*

**Abstract:** We experimentally demonstrate a novel scheme of power loading based on portfolio theory for millimeter-wave small-cell densification. By exploiting the statistical characteristics of interference, this approach improves the average throughput by 91% and reduces the variance.
**OCIS codes:** (060.5625) Radio frequency photonics; (060.2330) Fiber optics communications; (060.4256) Network optimization

## 1. Introduction

As one of the key performance attributes of 5G network, millimeter wave (MMW) wireless communication extends spectrum utilization and enhances system throughput [1,2]. MMW links exhibit more static and flat channel characteristics due to its directivity than conventional low-frequency wireless channels [3]. The high propagation loss makes it more suitable for short-distance transmission and small-cell deployment. Small-cell densification is one of the core challenges in 5G to enhance throughput and coverage of massive devices. One of the primary challenges in small-cell densification is inter-cell interference (ICI). Enhancing the overall throughput as well as maintaining the quality of service (QoS) under the impact of stochastic ICI has become an essential issue. Taking the advantage of the static and flat channel characteristics of MMW links, proper group action in power loading with respect to interference patterns becomes a desirable approach to address the issue [4,5].

Instead of the volatile channel conditions in conventional low-frequency wireless links, power loading in MMW small cells encounters interference patterns generated by signals from neighboring cells that have certain statistical properties. It is thus enticing to take the statistical information into account in the process of power allocation. In addition, 5G is a service-oriented network [6]. Applications such as vehicle-to-vehicle communications do not require ultra-high data rates but require guaranteed constant-rate transmission [7]. This contrasts with applications such as large file downloading, which require high data rates but allow for fluctuations in speed. The main idea behind power loading is to maximize the average throughput subject to a given level of uncertainty or risk.

In economics, Portfolio Theory (PT) is a well-known tool to develop investment strategies by optimizing a basket of assets based on mean-variance analysis [8,9]. Instead of solely focusing on maximizing the overall return, PT jointly assesses the return (mean) and risk (variance) of a portfolio. Like people who aim for high returns but controlling risk, PT-based power loading aims at enhancing the average throughput while controlling its variance, taking into account the statistical characteristics of the desired signal quality under the presence of ICI.

In this paper, power loading based on PT is proposed to optimize the throughput performance in a MMW small cell. Supported by analog radio-over-fiber (A-RoF) mobile fronthaul, 60-*GHz* MMW is used for the wireless links between the remote radio unit (RRU) and the user, as shown in Fig. 1. Due to the small-cell deployment in 5G, the user equipment (UE) receives interference generated by neighboring RRUs. With PT-based power loading among sub-bands, the power allocation (portfolio) providing optimal throughput is experimentally demonstrated.

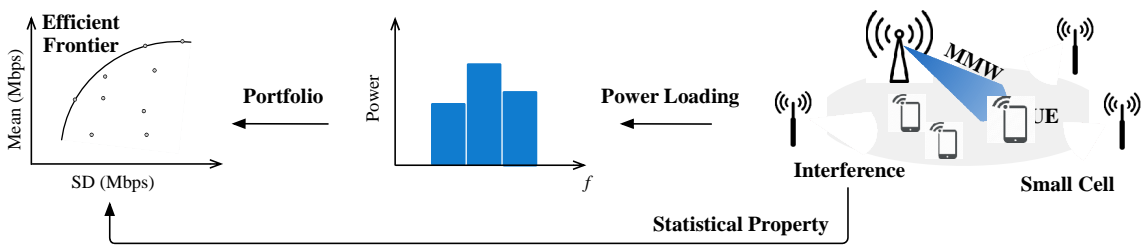

Fig. 1. Power loading based on PT for MMW small-cell densification.

## 2. Operating Principles

By holding a combination of assets and allocating different weights, PT generates mean-variance pairs of different portfolios which are bounded by the efficient frontier (EF) [8], as shown in Fig. 1. The portfolios on the EF provide maximum return for a given level of risk. Mapping to the case of a user accessing MMW frequencies, power allocation on *M* sub-bands comprises an *M*-asset portfolio. The mean and variance of the overall throughput correspond to those of the asset return. Accordingly, power loading based on PT yields an EF consisting of the optimal mean-variance pairs of the overall throughput. The EF serves as the guide to achieve the targeted average throughput with the given level of variance.

For *M* sub-bands, the throughput of each sub-band is $C_i$, $i=1,2,...,M$, and the variance is $\sigma_{C,i}^2$. For a given total power $P_{total}$, the power allocation portfolio is $\mathbf{P}$, where $\mathbf{P}^T = [p_1, p_2, ..., p_M]$, and $\sum_i p_i = P_{total}$. Given the power allocated by the portfolio, the signal-to-interference-plus-noise ratio (SINR) of the *i*th sub-band is $\gamma_i = \frac{p_i |H_i|^2}{I_i + 2\sigma_i^2}$, where $|H_i|^2$ is the channel gain, $I_i$ is the interference power, and $2\sigma_i^2$ is the noise power. The sub-band capacity is $C_i = W_i \log_2(1+\gamma_i)$, where $W_i$ is the sub-band bandwidth. The overall throughput is hence $C = \sum_i C_i$. The variance of the overall throughput is $\sigma_C^2 = \sum_i \sigma_{C,i}^2 + \sum_i \sum_{j \neq i} \sigma_{C,i} \sigma_{C,j} \rho_{i,j}$, where $\rho_{i,j}$ is the correlation coefficient.

According to PT, the problem is to maximize $E(C) = \sum_i E(C_i) = \sum_i E[W_i \log_2(1 + p_i \frac{|H_i|^2}{I_i + 2\sigma_i^2})]$, given $\sigma_C^2 = \sum_i \sigma_{C,i}^2(p_i) + \sum_i \sum_{j \neq i} \sigma_{C,i}(p_i) \sigma_{C,j}(p_j) \rho_{i,j}$, where $\sigma_{C,i}^2(p_i) = Var[W_i \log_2(1 + p_i \frac{|H_i|^2}{I_i + 2\sigma_i^2})]$, subject to $\sum_i p_i = P_{total}$. For all legitimate portfolios, the mean-variance pairs should be bounded by the EF, on which the optimal portfolios will reside and provide power allocation strategies.

### 3. Experimental Setup

The experimental setup is shown in Fig. 2. MMW signal is generated through optical-carrier-suppression (OCS) modulation. A light wave at 1549.8 nm is fed into a Mach-Zehnder modulator (MZM) biased at the null point. 30-*GHz* microwave generated by a radio-frequency (RF) source is fed into the MZM to obtain 60-*GHz* separation with optical-carrier suppression greater than 15-dB, as shown in Fig.2, inset (a). The desired OFDM signal and the interference signal generated by an arbitrary waveform generator (AWG) are modulated to the optical signals through two MZMs, respectively. The modulated optical signal is then amplified by an erbium-doped fiber amplifier (EDFA) and transmitted through a10-*km* standard single-mode fiber (SSMF). After detected by a photodetector (PD), the 60-*GHz* RF signal is amplified, then transmitted by a 15-dBi horn antenna (HA). Together with the interference signal, the overall signal is received by the receiver. Finally, the electrical signal is down-converted and captured by an oscilloscope followed by offline digital signal processing. The experimental results of the bit-error rate (BER) at RRU1 versus the received optical power are shown in Fig. 2, inset (b). It can be seen that the BER of RRU1 is significantly degraded when ICI exists.

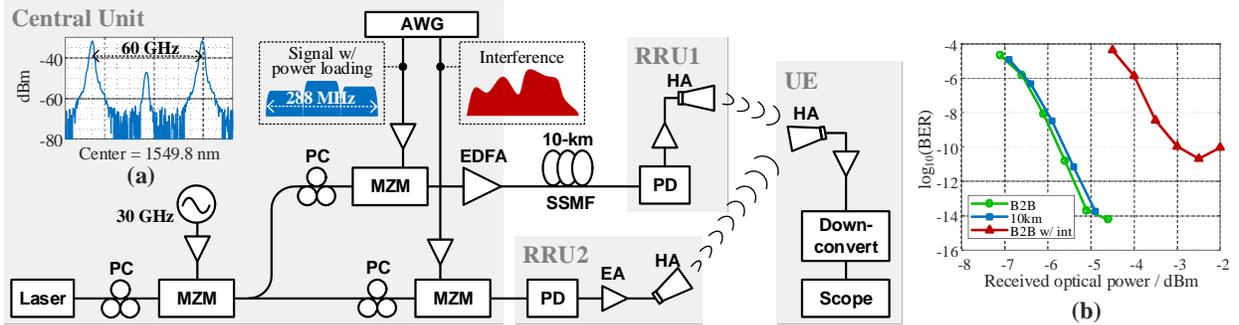

Fig. 2. Experimental setup. (a) Optical spectrum of optical MMW carrier; (b) BER vs. received optical power at UE.

In the experiment, the signal follows 5G orthogonal-frequency-division-multiplexing (OFDM) numerology [10]. The Fast-Fourier-Transform (FFT) size is 2048, 1200 of which are the payload. Subcarrier spacing is $15\ kHz \times 2^4 = 240\ kHz$, the total bandwidth is 288 *MHz*. The user is assigned one OFDM carrier by RRU1. In the experiment, the carrier is partitioned to *M* = 3 sub-bands, with sub-band bandwidth of 96 *MHz* consisting of 400 subcarriers. Accordingly, the portfolio **P** is a 3-by-1 vector, assigning power of $p_i$ to the *i*th sub-band. For the interference signal, three active interfering sub-bands with different mean-variance patterns are generated at RRU2. Power loading for three interference levels is investigated in the experiment.

### 4. Experimental Results

Experimental results and PT-generated portfolios under the three interference levels are shown in Fig. 3 (a), (b), and (c), respectively. To obtain the EF, 5000 randomly generated portfolios are plotted on each mean-standard deviation

(Mean-SD) plane. The EF is the upper left boundary of the region. Portfolios along the EF provides the maximum mean throughput for a given level of throughput variance. The optimal portfolio providing the overall maximum throughput for each case is marked as a green circle. These portfolios were experimentally examined, as shown by the blue squares, and match the results of the PT analysis. For the scenario shown in Fig. 3(a), the optimal portfolio is given by $\boldsymbol{P}_a^T = [0.35, 0.28, 0.37] \cdot P_{total}$, and provides the maximum throughput of 1350 *Mbps* with a SD of 342 *Mbps*. The cases where total power $P_{total}$ is allocated to a single sub-band are marked by red triangles. Compared with the single-sub-band portfolio that provides the highest throughput (705 Mbps), the PT-generated maximum-mean portfolio achieves 91% throughput improvement.

The portfolios that provide the minimum throughput variance are those when $P_{total}$ is allocated to the sub-band with the lowest variance while the rest two sub-bands are left idle. Along the EF from the portfolio with lowest variance to the portfolio with highest mean, the system can choose the power loading scheme customized for a targeted service requirement. Consequently, the EF generated by PT provides guide and flexibility of the power loading for service-oriented communications.

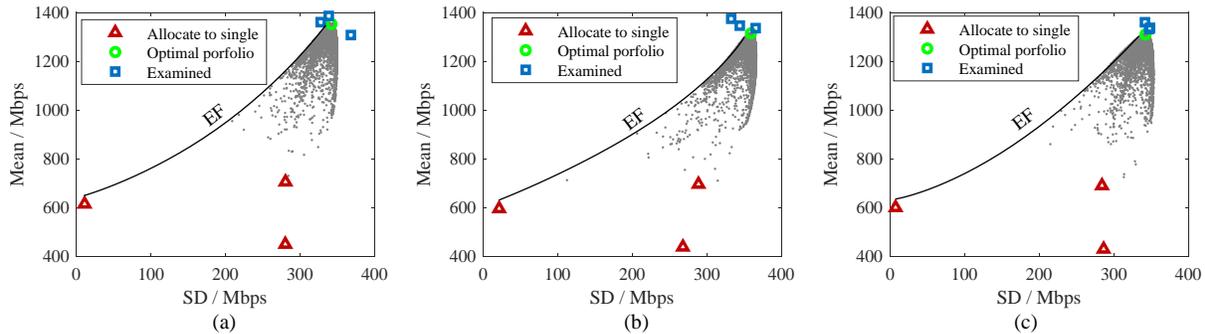

Fig. 3. Experimental results and portfolio results of PT analysis in mean-SD plane. Power level of the interference: (a) < (b) < (c).

## 5. Conclusion

We have demonstrated power loading based on portfolio theory (PT) for service-oriented communications in a MMW small-cell testbed. To cope with increasing ICI arising in densified small cells, power loading in MMW communication based on PT takes the statistical characteristics of interference signals into account and optimizes the overall throughput with respect to different interference pattern as well as performance targets of different applications. Optimal power loading portfolio based on PT analysis is experimentally demonstrated and verified. The portfolio with the maximum average throughput achieves 91% improvement compared to single-sub-band scenarios, providing an average throughput of 1350 *Mbps* with a controlled SD of 342 *Mbps*. In addition, with the EF given by PT analysis, the system is able to customize the mean and variance of the throughput optimized for various interference patterns, thus it enables service-oriented performance.